\documentclass[aps,prd,reprint]{revtex4-1}

\usepackage{dcolumn}
\usepackage{latexsym,array,theorem,mathrsfs,subfigure,bm,float}
\usepackage{amsmath}
\usepackage{amsfonts}
\usepackage{amssymb}
\usepackage{xcolor}
\usepackage[colorlinks=true, pdfstartview=FitV,linkcolor=blue, citecolor=blue, urlcolor=blue]{hyperref}

\newcommand{\be}{\begin{eqnarray}}
\newcommand{\ee}{\end{eqnarray}}
\newcommand{\nn}{\nonumber}
\newcommand{\muu}{M^2}
\newcommand{\mh}{m_{\rm H}}

\newcommand{\h}{{\cal X}}
\begin{document}

\title{Massive gravitons in arbitrary spacetimes}

\author{Charles Mazuet}
\email{\tt charles.mazuet@lmpt.univ-tours.fr}

\affiliation{
Laboratoire de Math\'{e}matiques et Physique Th\'{e}orique CNRS-UMR 7350, \\ 
Universit\'{e} de Tours, Parc de Grandmont, 37200 Tours, France}

\author{Mikhail~S.~Volkov}
\email{\tt volkov@lmpt.univ-tours.fr}
\affiliation{
Laboratoire de Math\'{e}matiques et Physique Th\'{e}orique CNRS-UMR 7350, \\ 
Universit\'{e} de Tours, Parc de Grandmont, 37200 Tours, France}

\affiliation{
Department of General Relativity and Gravitation, Institute of Physics,\\
Kazan Federal University, Kremlevskaya street 18, 420008 Kazan, Russia
}

\begin{abstract} 

We present two different versions of the consistent theory of massive gravitons in 
 arbitrary spacetimes which are simple enough for practical applications. 
 The theory is described by a non-symmetric  rank-2 tensor  whose equations of motion 
 imply six algebraic and five differential constraints reducing the number of independent components 
  to five.   The theory reproduces 
 the standard description of massive gravitons in Einstein spaces. 
 In generic spacetimes it does not show the massless limit and 
 always propagates five degrees of freedom, even for the vanishing mass parameter. 
 We illustrate these features by an explicit calculation for a homogeneous and isotropic 
 cosmological background. We find that the gravitons are stable if they are sufficiently massive, 
 hence they may be a part of Dark Matter at present. 
 We discuss also other possible applications.

\end{abstract} 

\maketitle

\section{Introduction}
Equations of massive fields of spin $0,1/2,1,3/2$ in Minkowski space (the Klein-Gordon, Dirac, Proca,
Rarita-Schwinger)  directly generalize to curved space, but for the massive spin 2
field this does not work.  The Fierz-Pauli (FP) theory of free massive 
gravitons  \cite{Fierz:1939ix}
generalises to curved space only for special  spacetimes -- Einstein spaces, whose Ricci tensor 
is proportional to the metric, $R_{\mu\nu}=\Lambda g_{\mu\nu}$ 
 \cite{Aragone:1971kh,*Aragone:1979bm,Higuchi}. In an arbitrary spacetime  the theory 
 shows six instead of five dynamical graviton polarizations, the extra polarization state being ghost-type.
 This feature 
 was for a longtime  thought  to be inevitable \cite{Buchbinder:1999ar}, hence 
 all applications of the consistent massive spin-2 theory have been limited 
 only to Einstein spaces. 
 Quite recently, by applying the 
 methods of the ghost-free massive gravity theory \cite{deRham:2010kj}, it was shown that 
 a consistent theory of massive gravitons can nevertheless  be formulated for arbitrary backgrounds
 \cite{Bernard:2014bfa,*Bernard:2015mkk,*Bernard:2015uic}.  However, the graviton mass term 
 obtained in \cite{Bernard:2014bfa,*Bernard:2015mkk,*Bernard:2015uic} is very complicated and 
 even the very demonstration of the existence of the constraint removing the sixth polarization requires 
 tedious calculations. 
 
 In what follows we present two different versions of the consistent theory of massive gravitons in 
 arbitrary spacetimes which are simple enough for practical applications. 
 The essential property of this theory is that it propagates only five and not six degrees of freedom 
 (DoF) and hence does not  show the extra polarization state. 
 In this sense the theory is ghost-free, which property is exceptional. As explained above, the 
 existence of such a theory for arbitrary backgrounds was  for a long time thought to be impossible. 
 Our theory is probably equivalent to that considered in \cite{Bernard:2014bfa,*Bernard:2015mkk,*Bernard:2015uic}
 since it is constructed in a similar way, but the equivalence is not immediately seen because our 
 parametrization is quite different.  Contrary to what is usually done, 
  we describe the massive spin-2 field by a  {\it non-symmetric} rank-2  tensor.
 Although this may seem odd, 
our  parameterization gives simple  results, which opens up 
 the possibility to efficiently study   massive gravitons in curved space. 
 
 We deliberately do not present details of our derivation -- they are rather technical and 
 will be given in a separate publication. Instead, we pass directly to the result: the theory described by 
 equations \eqref{0}--\eqref{2a} below.  It is easy to check that these equations propagate only 5 DoF in an 
 arbitrary spacetime, which is the key property. This property will be demonstrated in what follows 
by explicit calculations -- by counting the constraints for a generic background and also by explicitly solving the equations 
 for a cosmological background and counting the number of independent solutions. 
 In the latter case we show that the system is free of ghosts 
 and tachyons, hence massive gravitons are stable and could 
 contribute to the Dark Matter. We therefore have a valid new theory in our disposal 
 and we shall indicate some of its possible applications.

\section{Field equations}
Our theory of massive spin-2 field 
is parameterized  by a second rank tensor $X_{\mu\nu}$ which is apriory non-symmetric and 
has 16 independent components (one denotes $X=X^\sigma_{~\sigma}$). This field propagates 
in a spacetime with the standard (symmetric) Lorentzian metric $g_{\mu\nu}$ whose  Christoffel connection 
determines  the covariant derivative $\nabla_\mu$. 
The field equations are  
\be                         \label{0}
E_{\mu\nu}\equiv \Delta_{\mu\nu}+M_{\mu\nu}=0
\ee 
with 
\be                    \label{Del}
\Delta_{\mu\nu}&=&\frac12\nabla^\sigma\nabla_\mu (X_{\sigma\nu} +X_{\nu\sigma})
 +\frac12\nabla^\sigma\nabla_\nu (X_{\sigma\mu} +X_{\mu\sigma}) \nn \\
&-&\frac12\Box ({X}_{\mu\nu}+X_{\nu\mu})-\nabla_\mu\nabla_\nu{X}  \nn \\
 &+&g_{\mu\nu}\left(\Box X-\nabla^\alpha\nabla^\beta X_{\alpha\beta}+R^{\alpha\beta}X_{\alpha\beta}
\right) \nn  \\
& -&R^\sigma_\mu X_{\sigma\nu}-R^\sigma_\nu X_{\sigma\mu},
\ee
and we shall consider two different options for the mass term:
\be                     \label{2}
\mbox{\rm model I:}~~~~~
M_{\mu\nu}&= &\gamma_{\mu\alpha} X^\alpha_{~\nu}
-g_{\mu\nu}\,\gamma_{\alpha\beta}X^{\alpha\beta},  \nn \\
\gamma_{\mu\nu}&=&R_{\mu\nu}+\left(\muu -\frac{R}{6}\right)g_{\mu\nu}\,;
\ee
\be                   \label{2a}
\mbox{\rm model II:}~~~~M_{\mu\nu}&=& -X_\mu^{~\alpha} \gamma_{\alpha\nu} 
+X\gamma_{\mu\nu}\,,  \nn \\
\gamma_{\mu\nu}&=&R_{\mu\nu}-\left(\muu +\frac{R}{2}\right)g_{\mu\nu}\,.
\ee
Here $M$ is the FP  mass of gravitons. 
%
Since  $\Delta_{\mu\nu}$ is symmetric with respect to $\mu\leftrightarrow\nu$, the 
asymmetric part of the field equations, $E_{[\mu\nu]}=0$, yields algebraic conditions  
\be                       \label{3} 
\mbox{\rm I:}~~
\gamma_{\mu\alpha} X^\alpha_{~\nu}=\gamma_{\nu\alpha} X^\alpha_{~\mu}\,;~~~~
\mbox{\rm II:}~~X_\mu^{~\alpha} \gamma_{\alpha\nu} =X_\nu^{~\alpha} \gamma_{\alpha\mu} \,.
\ee
The equations \eqref{0} can be obtained by varying the action 
\be                        \label{act}
I=\frac12\int X^{\nu\mu}E_{\mu\nu}\sqrt{-g}\,d^4x\equiv \int L \,\sqrt{-g}\,d^4x
\ee
(notice the order of indices) 
where, after integrating by parts, the Lagrangian $L$ contains only the first derivatives of $X_{\mu\nu}$ 
(see the Appendix). One can directly check that 
 $\delta I=\int E_{\mu\nu}\,\delta X^{\nu\mu}\,\sqrt{-g}\,d^4x$. 

We shall only briefly indicate  how these theories were obtained (details will be given separately). 
The strategy  was to linearise equations of the ghost-free massive gravity \cite{deRham:2010kj}. 
When applied within the metric parameterization, this gives symmetric 
expressions for both the graviton kinetic 
and graviton mass terms, but the latter turns   out to be extremely complicated  
\cite{Bernard:2014bfa,*Bernard:2015mkk,*Bernard:2015uic}. 
We used  instead the tetrad parametrization.  
Perturbations of the background tetrad are  described in this approach by the non-symmetric $X_{\mu\nu}$ while 
perturbations of the background potential comprise the  mass term $M_{\mu\nu}$, which is apriori non-symmetric. 
However,
the kinetic term $\Delta_{\mu\nu}$ is symmetric since it is 
obtained by linearising the background Einstein tensor. 
This procedure actually gives many different theories,
but the above two models are the only ones for which the mass term 
$M_{\mu\nu}$ depends on  the background Ricci tensor {linearly}. 

We shall now check  that the number of propagating DoF in our models is indeed 5.

\section{Einstein spaces}
Let us consider first the special case where $R_{\mu\nu}=\Lambda g_{\mu\nu}$. Then, in both models,
the tensor $\gamma_{\mu\nu}$ becomes proportional to the metric and the conditions \eqref{3} yield 
$X_{\mu\nu}=X_{\nu\mu}$. The field equations reduce to 
\be             \label{4}
\Delta_{\mu\nu}+M_{\rm H}^2(X_{\mu\nu}-Xg_{\mu\nu})=0
\ee
where the Higuchi mass \cite{Higuchi} is 
$M_{\rm H}^2={\Lambda}/{3}+\muu$ in model {I} and 
$M^2_{\rm H}=\Lambda+\muu$ in model II. 
The operator $\Delta_{\mu\nu}$ is divergence free in this case (see the Appendix)
and is invariant under 
$X_{\mu\nu}\to X_{\mu\nu}+\nabla_{(\mu} \xi_{\nu)}$.
Therefore, for  $M_{\rm H}=0$ the theory \eqref{4} describes 
massless gravitons with two polarizations. For $M_{\rm H}\neq 0$, taking the divergence of \eqref{4}, one obtains 
four constraints $\nabla^\mu X_{\mu\nu}=\nabla_\nu X$ which reduce  the number of independent components 
of $X_{\mu\nu}$ to 6 and bring  equations \eqref{4} to the form
\be                \label{5}
&-&\Box X_{\mu\nu}+\nabla_\mu\nabla_\nu X-2R_{\mu\alpha\nu\beta}X^{\alpha\beta} \nn \\
&+&\Lambda Xg_{\mu\nu}
+M^2_{\rm H}(X_{\mu\nu}-Xg_{\mu\nu})=0.
\ee
Taking the trace of these yields 
$
(2\Lambda-3M_{\rm H}^2)X=0
$
hence, unless for $M^2_{\rm H}=2\Lambda/3$, one has $X=0$.
This  is the fifth constraint reducing the 
number of DoF to five. 
For $M^2_{\rm H}=2\Lambda/3\equiv M^2_{\rm PM}$ 
equations \eqref{5} 
admit the gauge symmetry 
$X_{\mu\nu}\to X_{\mu\nu}+\nabla_\mu\nabla_\nu\Omega+(\Lambda/3)g_{\mu\nu}\Omega$ 
and 
there remain only DoF
 (the partially massless (PM) case \cite{Higuchi}). 
As a result, our theory successfully reproduces the 
properties of massive gravitons in Einstein spaces. 

\section{Generic spacetime} 
In an arbitrary spacetime geometry
$X_{\mu\nu}$ has  no symmetries, but the six algebraic 
conditions \eqref{3} reduce the number of its independent components to ten.
Next,  one can see that 
there are in addition five differential constraints.

Consider first model I. 
Taking the divergence of $E_{\mu\nu}$, the third and second derivatives of $X_{\mu\nu}$ contained in 
$\nabla^\mu E_{\mu\nu}$ cancel, while the first derivatives  turn out to be all 
proportional to $\gamma_{\mu\nu}$ (see the Appendix). 
Multiplying by the inverse  $\tilde{\gamma}^{\mu\nu}$ (one has 
$\tilde{\gamma}^{\mu\alpha}\gamma_{\alpha\nu}=\delta^\mu_\nu$ while the indices are moved by the metric,
so that $\gamma_{\mu\nu}=g_{\mu\alpha}\gamma^\alpha_{~\nu}$
and $\tilde{\gamma}^\mu_{~\nu}=g^{\mu\alpha}\tilde{\gamma}_{\alpha\nu}$) 
one obtains  the four vector constraints 
\be                  \label{vI}
{\rm I:}~~~
{\cal C}^\rho\equiv \tilde{\gamma}^{\rho\nu}\nabla^\mu E_{\mu\nu}=\nabla_\sigma X^{\sigma\rho}-\nabla^\rho X
+{\cal I}^\rho=0
\ee
with ${\cal I}^\rho= \tilde{\gamma}^{\rho\nu}\left\{X^{\alpha\beta}(\nabla_\alpha G_{\beta\nu}
-\nabla_\nu\gamma_{\alpha\beta})+\nabla^\mu\gamma_{\mu\alpha}X^\alpha_{~\nu}\right\}$. 
Taking the divergence of this and combining with the trace of the equations, the second derivatives cancel 
 yielding  the fifth constraint 
\be                \label{sI}
{\rm I:}~~{\cal C}_5&\equiv& \nabla_\rho {\cal C}^\rho+\frac12 E^\mu_{~\mu}  \nn \\
&=&-\frac32\muu X-\frac12 G^{\mu\nu}X_{\mu\nu}
+\nabla_\rho{\cal I}^\rho=0.~~~
\ee
Therefore, the number of propagating DoF is  $10-5=5$. 
If $R_{\mu\nu}=\Lambda g_{\mu\nu}$ then ${\cal I}^\rho=0$ and the constraints reduce to 
the same as before: $\nabla_\sigma X^{\sigma\rho}-\nabla^\rho X=0$ and 
$(\Lambda/3-\muu)X=(M^2_{\rm PM}-M^2_{\rm H})X=0$. 

Consider now model II. Taking the divergence of $E_{\mu\nu}$
and multiplying by $\gamma^{\rho\nu}$ (and not by $\tilde{\gamma}^{\rho\nu}$) yields 
\be                          \label{vII}
{\rm II:}~~{\cal C}^\rho\equiv \gamma^{\rho\nu}\nabla^\mu E_{\mu\nu}=
\Sigma^{\rho\nu\alpha\beta}\nabla_\nu\, X_{\alpha\beta}=0
\ee
with $\Sigma^{\rho\nu\alpha\beta}\equiv \gamma^{\rho\nu}\gamma^{\alpha\beta}-
\gamma^{\rho\beta}\gamma^{\nu\alpha}$. 
Taking the divergence again and combining with the 
field equations yields  (see the Appendix)
\be                         \label{sII}
{\rm II:}~~&&{\cal C}_5\equiv \nabla_\rho {\cal C}^\rho \\
&+&
\frac{1}{2g^{00}}\,\Sigma^{00\alpha}_{~~~~\beta}\left(2E^\beta_{~\alpha}
-\delta_\alpha^\beta\, (E^\sigma_{~\sigma}-\frac{1}{g^{00}}\,E^{00}  )\right)=0.  \nn
\ee
A remark is in order here. This expression does in general contain second derivatives of $X_{\mu\nu}$,
but  not the second time derivatives. The expression is not generally covariant and depends 
on  the choice of time, but for any such a choice the second derivatives with respect to the 
corresponding time coordinate drop out from ${\cal C}_5$. 
 Therefore, ${\cal C}_5=0$ is a constraint  restricting  the initial data.

Summarizing, the six algebraic conditions  \eqref{3} and five differential constraints ${\cal C}^\rho=0$ 
and ${\cal C}_5=0$ reduce the number of independent components of $X_{\mu\nu}$ from 16 to 5, which matches
the number of polarizations  of massive particles of spin 2. When restricted to Einstein spaces, 
our theory reproduces the standard description of massive gravitons. 
However, its unusual feature  is that, unless in Einstein spaces, 
the theory  does not show the massless limit, since the mass term $M_{\mu\nu}$ never vanishes 
for generic backgrounds, whatever the value of the mass parameter $M$ is. 
Therefore, unless in Einstein spaces,  the theory always propagated five DoF. 
We shall confirm this below also by explicitly solving the equations and counting 
the independent modes  in the general solution.

\section{Cosmological background}
As an application and 
in order to illustrate the above features, we explicitly construct the general solution of $E_{\mu\nu}=0$ on a 
cosmological background with 
$g_{\mu\nu}dx^\mu dx^\nu=-dt^2+a^2(t)d{\bf x}^2$
where $a(t)$ fulfills the Einstein equations 
\be                     \label{cosm}
3\,\frac{\dot{a}^2}{a^2}=\frac{\bm{\rho}}{M^2_{\rm Pl}}\equiv \rho,
~~~~2\,\frac{\ddot{a}}{a}+\frac{\dot{a}^2}{a^2}=-\frac{\bm{p}}{M^2_{\rm Pl}}\equiv -p.
\ee
Here $M_{\rm Pl}$ is the Planck mass and $\bm{\rho},\bm{p}$ are 
the energy density and pressure of the background matter. 
The general solution can be represented as 
$X_{\mu\nu}(t,{\bf x})=a^2(t)\sum_{\bf k}X_{\mu\nu}(t,{\bf k})e^{i{\bf kx}}$ where the Fourier amplitude 
splits into the sum of the tensor, vector, and scalar harmonics:
$X_{\mu\nu}(t,{\bf k})=X^{(2)}_{\mu\nu}+X^{(1)}_{\mu\nu}+X^{(0)}_{\mu\nu}
$. 
Since the spatial part of the background 
Ricci tensor is proportional to the unit matrix, $R_{ik}\sim \delta_{ik}$, the algebraic 
constraints \eqref{3} imply that $X_{ik}=X_{ki}$ hence $X_{\mu\nu}$ has in this case only 13 
independent components. Assuming the spatial momentum $\bf{k}$ to be directed along 
the third axis, ${\bf k}=(0,0,{\rm k})$, the harmonics can be parameterized as 
\begin{widetext}
\be                          \label{XX}
X^{(2)}_{\mu\nu}=
\begin{bmatrix}
    0       & 0 & 0 & 0 \\
    0       & {\rm D_+} & {\rm D_-} & 0 \\
    0       & {\rm D_-} & {\rm -D_+} & 0 \\
       0       & 0 & 0 & 0 \\
 \end{bmatrix},~
 X^{(1)}_{\mu\nu}=
\begin{bmatrix}
    0       & W_+^+ & W_-^+ & 0 \\
    W^-_+       & 0 & 0 & i{\rm k}{\rm V_+} \\
    W^-_-       & 0 & 0 & i{\rm k}{\rm V_-} \\
    0       & i{\rm k}{\rm V_+}& i{\rm k}{\rm V_-} & 0 \\
 \end{bmatrix},~
 X^{(0)}_{\mu\nu}=
\begin{bmatrix}
    S^+_+       & 0 & 0 & i{\rm k}S^+_- \\
    0       & S^-_- & 0 & 0 \\
    0       & 0 & S^-_- & 0 \\
   i{\rm k} S^-_+       & 0 & 0 & S^-_--{\rm k}^2{\rm S} \\
 \end{bmatrix},~~
\ee
\end{widetext}
where ${\rm D_\pm}$, ${\rm V}_{\pm}$, ${\rm S}$, $W^\pm_\pm$, $S^\pm_\pm$  are 
functions of time. Injecting everything to $E_{\mu\nu}=0$, 
the equations split into three independent groups -- one for the tensor modes 
$X^{(2)}_{\mu\nu}$, one for  vector modes $X^{(1)}_{\mu\nu}$, and one for  scalar modes 
$X^{(0)}_{\mu\nu}$. 

In the tensor sector everything reduces to two decoupled second order equations 
for ${\rm D}_+$ and for ${\rm D}_-$ describing  the two tensor polarizations. 
In the vector sector  the four amplitudes $W^\pm_\pm$ can be expressed (see the Appendix) in terms of 
${\rm V}_{+}$ and ${\rm V}_{-}$
which  fulfill two independent second order equations describing  the two vector polarizations. 

Most importantly, one finds that in the scalar sector 
the four $S^\pm_\pm$ can be expressed in terms of one single amplitude S that fulfills a second order 
master equation (see the Appendix). 
Therefore, there is only one scalar polarization, hence the total number of polarizations is 5. 

Injecting everything to the action \eqref{act}, it splits into the sum of five independent 
terms of the form 
\be                    \label{act1}
\int (K \dot{Y}^2-UY^2)\, a^3 dt. 
\ee
Varying this yields the master equation in each sector. 
For the tensor modes one has 
$Y=D_{+}$ or $Y=D_{-}$ and $K=1$ while $U=M^2_{\rm eff}+{\rm k}^2/a^2$.
Here and in what follows  we denote, depending on the model,
\be                      \label{Meff}
{\rm I}:~~~M^2_{\rm eff}&=&M^2+\frac13\,\rho,~~~~\mh^2=M^2_{\rm eff},  \nonumber \\
{\rm II}:~~~M^2_{\rm eff}&=&M^2-p,~~~~\mh^2=M^2+\rho, 
\ee
where $M_{\rm eff}$ is the effective graviton mass  (notice that $M^2_{\rm eff}$ may be negative) while 
$\mh$ reduces to the Higuchi mass in the Einstein  space limit. 

For the vector modes one has $Y={\rm V}_{+}$ or $Y={\rm V}_{-}$ and 
\be                               \label{Kv}
K\equiv K_{(1)}=\frac{{\rm k}^2 m^4_{\rm H}}{m^4_{\rm H}+({\rm k}^2/a^2)(m_{\rm H}^2-\epsilon/2)}
\ee
where $\epsilon=\rho+p$ while the potential is $U=M^2_{\rm eff}\, {\rm k}^2$. 

In the scalar sector one has $Y={\rm S}$ and, introducing the Hubble parameter $H=\dot{a}/a$, 
the kinetic term is 
\begin{widetext}
\be                              \label{Ks}
K\equiv K_{(0)}=\frac{3{\rm k}^4 m^4_{\rm H}(m_{\rm H}^2-2H^2)}
{(m_{\rm H}^2-2H^2)[9m_{\rm H}^4+6({\rm k}^2/a^2) (2m_{\rm H}^2-\epsilon )]
+4({\rm k}^4/a^4)(m_{\rm H}^2-\epsilon)
}.
\ee
\end{widetext}
The  potential in the scalar sector  is more complicated (see the Appendix) but its asymptotic behaviour is simple.
In all sectors one has $U/K\to M_{\rm eff}^2$ for ${\rm k}\to 0$  while  for ${\rm k}\to \infty $ one has
$U/K\to c^2\,({\rm k}^2/a^2)$ where $c$ is the sound speed. 
One has for the vector and scalar  modes, respectively,  
\be       \label{cs}
c_{(1)}^2&=&\frac{M_{\rm eff}^2}{\mh^4}\,(\mh^2-\epsilon/2),   \\
c_{(0)}^2&=&\frac{(\mh^2-\epsilon)[\mh^4+(2H^2-4M^2_{\rm eff}-\epsilon)\mh^2+4H^2M^2_{\rm eff} ]   }
{3\mh^4(2H^2-\mh^2)},  \nonumber 
\ee
while  for the tensor modes $c^2_{(2)}=1$. For the vectors and scalars one has 
$c^2<1$ but $c^2\to 1$  if $\rho\to 0$. 

If  $R_{\mu\nu}=\Lambda g_{\mu\nu}$  then $\rho=-p=\Lambda$ 
while 
  $m^2_{\rm H}= M^2_{\rm H}$  and $2H^2= M^2_{\rm PM}=2\Lambda/3$. 
The above formulas  then imply that 
if $M_{\rm H}=0$ then $K_{(0)}=K_{(1)}=0$, 
hence  only  the tensor modes propagate. 
The massless theory is recovered in this way. If $0<M_{\rm H}<M_{\rm PM}$ 
then $K_{(0)}<0$ (for $k\to\infty$)  and the scalar polarization becomes 
(Higuchi) ghost \cite{Higuchi}. If $M_{\rm H}=M_{\rm PM}$ then $K_{(0)}=0$
and  the scalar polarization 
 is non-dynamical (the PM case). 

All these features  are well known for massive gravitons in Einstein spaces. 
However,  for generic backgrounds, where $\rho,p$ are not constant, 
$m_{\rm H}$ and 
$H$ become functions of time, and it is not possible 
to have $K_{(1)}=0$ or $K_{(0)}=0$ for all time moments, whatever the value of the FP mass $M$ is. 
Therefore, neither the massless nor PM  cases are contained in the theory
on generic backgrounds and it always propagates  five polarizations. 
At most, there could be special backgrounds 
where gravitons become massless or PM for some values of $M$ \cite{Bernard:2017tcg}.

A direct inspection of Eqs.\eqref{Kv}--\eqref{cs} shows that 
if $\rho$ is small, 
$\rho\leq M^2$ 
 ($\bm{\rho}\leq M^2 M_{\rm Pl}^2$), then 
$K>0$ (for ${\rm k}\to\infty$) and $c^2>0$, 
hence the system is free of ghosts and tachyons. 
{The situation is more complex for large $\rho$. 
In model I  the kinetic term $K_{(0)}$  changes sign  for $\rho>3M^2$
 since $\mh^2<2H^2$ in this case, 
which corresponds to the Higuchi ghost. However, $c_{(0)}^2$ also 
changes sign at the same time, unless for $p/\rho=-1$,  so that  
the ghost and tachyon ``compensate each other" only changing 
the overall sign of the action. 
In model II  one always has  $\mh^2>2H^2$ and 
 the Higuchi ghost is absent, but since $M^2_{\rm eff}$ may be negative, 
 there could be tachyons  in the vector sector. 
 However, one finds in this case 
  that $K>0$ (always for ${\rm k}\to\infty$) and that $c^2>0$
for any $\rho$ provided that $p/\rho <-2/5$. 
Therefore, model II is  stable during the 
 inflationary stage, whereas model I is stable if the graviton mass is large enough, $M\geq H$. 
 Estimating that $\bm{\rho}\approx (10^{16}{\rm GeV})^4$ 
 at the beginning of the radiation dominated stage \cite{weinberg}, 
 it follows that for $M\geq 10^{13}$ GeV one would have 
 ${\rho} \leq M^2$ and hence both models I and  II would  be stable at all times after the inflation.

A much milder bound $M\geq 10^{-3}$ eV is needed to insure that both models  are stable 
at present when $\rho$ is small. 
Assuming that the tensor $X_{\mu\nu}$ couples only to the gravity 
and hence massive gravitons do not have other decay channels, 
it follows that they could be a part of Dark Matter (DM) at present. 
 Massive gravitons as DM candidates have actually been considered  before 
  \cite{Dubovsky:2004ud,*Aoki:2016zgp,*Babichev:2016bxi,*Aoki:2017cnz},
 but only  our description  is consistent for arbitrary  backgrounds.

  One may wonder if the massive gravitons could also mimic the dark energy as in the 
  other massive gravity models \cite{Volkov:2013roa}. 
  To calculate their backreaction on the spacetime geometry, one adds 
  the Einstein-Hilbert term to the action  \eqref{act}, which yields 
  \be                           \label{I}
  I=\frac12\int\left(M_{\rm Pl}^2R+X^{\nu\mu}E_{\mu\nu}\right)\sqrt{-g}\, d^4x.
  \ee
  Varying this with respect to the metric yields Einstein equations 
  $M_{\rm Pl}^2 G_{\mu\nu}=T_{\mu\nu}$ 
  to be solved 
  together with $E_{\mu\nu}=0$. The energy-momentum tensor 
  $T_{\mu\nu}$ has a somewhat complicated structure,
  partly  due to the non-minimal terms like $X^{\mu\nu}R^\sigma_{\nu}X_{\sigma\mu}$ 
  in the action (see the Appendix). We solved the equations in the homogeneous and isotropic sector, 
  with  $X_{\mu\nu}=X^{(0)}_{\mu\nu}$ given by \eqref{XX} for ${\rm k}=0$, and 
 we found the  solution only in model II and only for $M^2<0$: 
  this is the de Sitter space with $\Lambda=-3M^2>0$. This would be a legitimate solution if we
  flipped  the sign in front of $M^2$ in \eqref{3} from the very beginning, 
  but such a modified theory would be 
  very unstable since $K$ and $c^2$ in \eqref{Kv}--\eqref{cs} would then be negative. 
  We therefore conclude that our theory  cannot mimic  a positive $\Lambda$-term. 
  
 Since the potential $U$ can be negative, one might wonder 
 if the gravitons could exhibit features like superluminality \cite{Deser:2012qx}. 
 We have seen that for the cosmological background this  problem does not arise, 
 since the system is free form ghosts and tachyons. 
 Other backgrounds should be studied separately.

  \section{Other possible applications}  
  Apart from cosmology, the theory 
  of massive spin 2 field can have other applications. 
  For example, massive gravitons in curved space can be used for 
  the holographic description of superconductors \cite{Benini:2010pr} 
  or electron-phonon interactions \cite{Baggioli:2014roa}.
  Up to now all applications have always been restricted to the Einstein spaces, 
  but in our theory this is no longer necessary.

  One can also study solutions with back-reacting  massive gravitons described by \eqref{I}, 
  as for example static stars  or black holes. 
  Yet more interesting applications could  be found by extending the theory 
   to complex values of  $X_{\mu\nu}$ via replacing 
  in \eqref{I} $X^{\nu\mu}E_{\mu\nu}\to \bar{X}^{\nu\mu}E_{\mu\nu}+X^{\nu\mu}\bar{E}_{\mu\nu}$ with the 
  bar denoting complex conjugation.  Very recently it was shown \cite{East:2017ovw}
 that the superradiance \cite{Bardeen:1972fi,*Starobinsky:1973aij} 
 of complex massive fields in the vicinity of spinning black holes 
leads to a spontaneous formation of massive clouds  evolving towards 
stationary hairy black holes \cite{Herdeiro:2014goa}. 
So far this phenomenon has been observed only 
for the complex Proca field, but since the superradiance rate increases rapidly with spin  
\cite{Bardeen:1972fi,*Starobinsky:1973aij}, it would be interesting 
to carry out a similar analysis   for the spin 2 complex field. 

\section{Summary} 
We have constructed a new theory of a free massive spin-2 field 
in curved space. This theory is exceptional because it propagates not more than 5 DoF  for any background. 
So far only one theory with this property has been discovered 
\cite{Bernard:2014bfa,*Bernard:2015mkk,*Bernard:2015uic}, to which theory our theory 
is probably equivalent, but we use a completely different parametrization in terms of a 
non-symmetric tensor $X_{\mu\nu}$, which yield simpler results. Our main goal 
was to show that our theory is self-consistent and that the number of independent DoF is indeed 5. 
We have shown this by counting the constraints and also by counting the independent modes in the 
general solution. It turns out that massive gravitons in the expanding universe are free of ghosts and tachyons. 

We notice finally that the recent LIGO data  \cite{Abbott:2016blz} imply 
 that the graviton mass should be less than $10^{-22}$ eV
 \cite{deRham:2016nuf}.  However, this  bound 
applies rather to the mass of quanta of the background metric $g_{\mu\nu}$ and not to that of $X_{\mu\nu}$. 
 It is consistent to assume that $X_{\mu\nu}$ 
 does not directly interact  with the ordinary matter and hence is not seen by the  LIGO detector. 
Therefore  the bound does not apply to the FP mass $M$.

 \section{Acknowledgments} 
 We thank Arkady Tseytlin and Matteo Beccaria
 for  critical remarks and confirming our formula \eqref{act} for the action. 
 We also thank 
Eugen Radu and Carlos Herdeiro  for explaining to us the 
 modern aspects of the 
 superradiance phenomenon, as well as Shinji Mukohyama for his critical remarks. 
 M.S.V. was partly supported by the Russian Government Program of Competitive Growth 
of the Kazan Federal University.

\section{Appendix}

Below we sketch some technical details 
of the calculations presented in the main text above.

\subsection{A. Lagrangian}

\setcounter{section}{0}
\setcounter{equation}{0}
\setcounter{subsection}{0}

\renewcommand{\theequation}{A.\arabic{equation}}

The Lagrangian $L$ in the action $I=\int L\sqrt{-g}\, d^4 x$
defined by Eq.\eqref{act} in the main text is $L=L_{(2)}+L_{(0)}$ where
\be
L_{(2)}=&-&\frac14\, \nabla^\sigma \h^{\mu\nu}\nabla_\mu \h_{\nu\sigma}
+\frac18\,\nabla^\alpha \h^{\mu\nu} \nabla_\alpha \h_{\mu\nu}   \nn \\
&+&\frac14 \nabla^\alpha \h \nabla^\beta \h_{\alpha\beta}-\frac18\, \nabla_\alpha \h\nabla^\alpha \h
\ee
with $\h_{\mu\nu}=X_{\mu\nu}+X_{\nu\mu}$ and $\h=\h^\alpha_{~\alpha}$. One has in model~I
\be
L_{(0)}=&-&\frac12\,X^{\mu\nu}R^\sigma_{~\mu}X_{\sigma\nu} \nn \\
&+&\frac12\,(M^2-\frac{R}{6})(X_{\mu\nu}X^{\nu\mu}-X^2)
\ee
and in model II
\be
L_{(0)}=&-&\frac12\,X^{\mu\nu}R^\sigma_{~\mu}X_{\sigma\nu} 
-\frac12\,X^{\mu\nu}R^\sigma_{~\nu}X_{\sigma\mu}   \nn \\
&-&\frac12X^{\mu\nu}X_{\nu\alpha}R^\alpha_{~\mu}+XR_{\mu\nu}X^{\mu\nu} \nn \\
&+&\frac12\,(M^2+\frac{R}{2})(X_{\mu\nu}X^{\nu\mu}-X^2);
\ee
the order of indices being important. One can directly check that 
varying the action with respect to $X_{\mu\nu}$ yields the field equations, 
\be
\delta I=\int E_{\nu\mu}\, \delta X^{\mu\nu}\,\sqrt{-g}\,d^4x.
\ee
Varying with respect to the metric gives the energy-momentum tensor, 
$\delta I=-2\int T_{\mu\nu}\,\delta g^{\mu\nu}\,\sqrt{-g}\,d^4 x$.

\subsection{B. Constraints }

\setcounter{section}{0}
\setcounter{equation}{0}
\setcounter{subsection}{0}

\renewcommand{\theequation}{B.\arabic{equation}}

Here we sketch the derivation of the five 
constraints expressed  by Eqs.\eqref{vI}--\eqref{sII} in the main text. 
Using 
$$(\nabla_\mu\nabla_\nu-\nabla_\nu\nabla_\mu)X^\alpha_{~\beta}
=R^\sigma_{~\beta\nu\mu}X^\alpha_{~\sigma}
-R^\alpha_{~\sigma\nu\mu}X^\sigma_{~\beta}$$
a direct calculation yields the following result for the divergence 
of $\Delta_{\mu\nu}$ defined by Eq.\eqref{Del} in the main text:
\be              \label{20}
\nabla^\mu\Delta_{\mu\nu}&=&
\bm{\gamma}_{\nu\beta}(\nabla_\alpha X^{\alpha\beta}-\nabla^\beta X)\, \nn \\
&+&\bm{\gamma}_{\alpha\beta}(\nabla_\nu X^{\alpha\beta}-\nabla^\alpha X^\beta_{~\nu})  \nn \\
&+&X^{\alpha\beta}\nabla_\alpha G_{\beta\nu}.
\ee
Here we introduced the bold-faced $\bm{\gamma}_{\alpha\beta}\equiv R_{\alpha\beta}+\bm{\phi}\, g_{\alpha\beta}$ 
where $\bm{\phi}$ 
can be set to any value, because the part of $\bm{\gamma}_{\alpha\beta}$ proportional to $g_{\mu\nu}$ 
cancels in \eqref{20}. If $R_{\mu\nu}=\Lambda g_{\mu\nu}$ then $\bm{\gamma}_{\alpha\beta}\sim g_{\mu\nu}$ 
and \eqref{20} yields $\nabla^\mu\Delta_{\mu\nu}=0$. 

The divergence of $M_{\mu\nu}$ in model I given by   Eq.\eqref{2} in the main text is 
\be              \label{21} 
\nabla^\mu M_{\mu\nu}=\gamma_{\alpha\beta}(\nabla^\alpha X^\beta_{~~\nu}-\nabla_\nu X^{\alpha\beta})  \nn \\
+X^\alpha_{~~\nu}\nabla^\mu \gamma_{\mu\alpha}-X^{\alpha\beta}\nabla_\nu \gamma_{\alpha\beta}
\ee
with $\gamma_{\alpha\beta}$ defined by Eq.\eqref{2}. Setting in \eqref{20} 
$\bm{\gamma}_{\alpha\beta}={\gamma}_{\alpha\beta}$ and adding up with \eqref{21},
the second line on the right in \eqref{20} cancels against the first line in \eqref{21}, yielding 
\be
\nabla^\mu(\Delta_{\mu\nu}+M_{\mu\nu})={\gamma}_{\nu\beta}(\nabla_\alpha X^{\alpha\beta}-\nabla^\beta X) \nn \\
+X^{\alpha\beta}\nabla_\alpha G_{\beta\nu}+
X^\alpha_{~~\nu}\nabla^\mu \gamma_{\mu\alpha}-X^{\alpha\beta}\nabla_\nu \gamma_{\alpha\beta}.
\ee
Multiplying this by the inverse $\tilde{\gamma}^{\rho\nu}$ of $\gamma_{\alpha\beta}$ one obtains
\be
{\cal C}^\rho\equiv \tilde{\gamma}^{\rho\nu}\nabla^\mu E_{\mu\nu}
=\nabla_\alpha X^{\alpha\rho}-\nabla^\rho X ~~~~~~~~~~~~ \\
+\tilde{\gamma}^{\rho\nu}(X^{\alpha\beta}\nabla_\alpha G_{\beta\nu}+
X^\alpha_{~~\nu}\nabla^\mu \gamma_{\mu\alpha}-X^{\alpha\beta}\nabla_\nu \gamma_{\alpha\beta}), \nn
\ee
which yields  Eq.\eqref{sI} in the main text. 
Acting on this with $\nabla_\rho$ and combining with the trace $E^\mu_{~\mu}$ reproduces 
Eq.\eqref{sI} in the main text. 

The divergence of $M_{\mu\nu}$ in model II given by  Eq.\eqref{2a} in the main text is 
\be              \label{22} 
\nabla^\mu M_{\mu\nu}=\gamma_{\nu\beta}(\nabla^\beta X-\nabla_\alpha X^{\alpha\beta}) \nn \\
+X\nabla^\mu\gamma_{\mu\nu}-X^{\alpha\beta}\nabla_\alpha\gamma_{\beta\nu}
\ee
with $\gamma_{\alpha\beta}=G_{\mu\nu}-M^2g_{\mu\nu}$. Setting in \eqref{20} 
$\bm{\gamma}_{\alpha\beta}={\gamma}_{\alpha\beta}$ and adding up with \eqref{22}, the first and third 
lines on the right in \eqref{20} cancel against \eqref{22}, hence 
\be              \label{20a}
\nabla^\mu(\Delta_{\mu\nu}+M_{\mu\nu})=
{\gamma}_{\alpha\beta}(\nabla_\nu X^{\alpha\beta}-\nabla^\alpha X^\beta_{~\nu}). 
\ee
Multiplying this by $\gamma^{\rho\nu}$  yields 
\be
{\cal C}^\rho&\equiv& \gamma^{\rho\nu}\nabla^\mu E_{\mu\nu}=\gamma^{\rho\nu}{\gamma}_{\alpha\beta}(\nabla_\nu X^{\alpha\beta}-\nabla^\alpha X^\beta_{~\nu}) \nn \\
&=&(\gamma^{\rho\nu}\gamma^{\alpha\beta}-\gamma^{\rho\beta}\gamma^{\nu\alpha})\nabla_\nu X_{\alpha\beta}\,,
\ee
which reproduces Eq.\eqref{vII} in the main text. Acting with $\nabla_\rho$ gives 
\be
\nabla_\rho{\cal C}^\rho&=&(\gamma^{\rho\nu}\gamma^{\alpha\beta}-\gamma^{\rho\beta}\gamma^{\nu\alpha})
\nabla_\rho\nabla_\nu X_{\alpha\beta}  \nn \\
&+&\nabla_\rho(\gamma^{\rho\nu}\gamma^{\alpha\beta}-\gamma^{\rho\beta}\gamma^{\nu\alpha})
\nabla_\nu X_{\alpha\beta} \nn \\
&=&(\gamma^{00}\gamma^{\alpha\beta}-\gamma^{0\beta}\gamma^{0\alpha})\ddot{X}_{\alpha\beta}+\ldots \\ \nn
&=&(\gamma^{00}\gamma^{ik}-\gamma^{0i}\gamma^{0k})\ddot{X}_{ik}+\ldots 
\ee
where 
the dots denote   terms not containing second time derivatives of $X_{\alpha\beta}$.
The definition of $\Delta_{\mu\nu}$ in Eq.\eqref{Del} in the main text implies that (here $i,k,m,n=1,2,3$)
\be
\Delta_{ik}=-g^{00}\ddot{X}_{(ik)}+g_{ik} \,g^{00} \,h^{nm} \ddot{X}_{nm}+\ldots 
\ee
with $h^{ik}=g^{ik}-g^{0i}g^{0k}/g^{00}$ hence 
\be
\ddot{X}_{(ik)}&=&\frac{1}{g^{00}}\left(\frac{1}{2}\,{g_{ik}}\,h^{nm}\Delta_{nm}-\Delta_{ik}  \right)+\ldots \nn \\
 &=&\frac{1}{g^{00}}\left(\frac{1}{2}\,{g_{ik}}\,h^{nm} E_{nm}-E_{ik}  \right)+\ldots
\ee
It follows that the combination 
\be
\frac{1}{g^{00}}(\gamma^{00}\gamma^{ik}-\gamma^{0i}\gamma^{0k})\left(\frac{1}{2}\,{g_{ik}}\,h^{nm} E_{nm}-E_{ik}  \right)
~~~~~~ \\
=\frac{1}{g^{00}}(\gamma^{00}\gamma^{\alpha\beta}
-\gamma^{0\alpha}\gamma^{0\beta})\left(\frac{1}{2}\,{g_{\alpha\beta}}\,h^{nm} E_{nm}-E_{\alpha\beta}  \right) \nn
\ee
has precisely the same second time derivatives as in $\nabla_\rho{\cal C}^\rho$. Noting finally that 
\be
h^{nm} E_{nm}&=&\frac{1}{g^{00}}(g^{00}g^{mn}-g^{0m}g^{0n}) E_{mn}   \\
&=&\frac{1}{g^{00}}(g^{00}g^{\mu\nu}-g^{0\mu}g^{0\nu}) E_{\mu\nu}
= E^\alpha_{~\alpha}-\frac{1}{g^{00}}\, E^{00}  \nn
\ee
 yields Eq.\eqref{sII}  in the main text.

\subsection{C. Solution in the expanding universe}

\setcounter{section}{0}
\setcounter{equation}{0}
\setcounter{subsection}{0}

\renewcommand{\theequation}{C.\arabic{equation}}

Injecting Eqs.\eqref{cosm},\eqref{XX} from the main text to the equations $E_{\mu\nu}=0$ expressed by Eq.\eqref{0}, 
the direct inspection reveals that the vector amplitudes $W^\pm_{\pm}$ in Eq.\eqref{XX} can be 
expressed in terms of two independent amplitudes ${\rm V}_\pm$ as
\be
W^{+}_\pm =\frac{{\rm P}^2 m_{\rm H}^2\, \dot{\rm V}_\pm}{
m_{\rm H}^4+{\rm P}^2(m_{\rm H}^2-\epsilon/2)
},    \nonumber \\
W^{-}_\pm =\frac{{\rm P}^2\,[m_{\rm H}^2-\epsilon]\, \dot{\rm V}_\pm}{
m_{\rm H}^4+{\rm P}^2(m_{\rm H}^2-\epsilon/2)
},   \nonumber \\
\ee 
where  $m_{\rm H}$ and $\epsilon$ are defined after Eq.\eqref{Meff} in the main text and ${\rm P}={\rm k}/a$ 
is the physical momentum.  The equations 
for ${\rm V}_\pm$ reduce to those obtained by varying the effective action 
expressed by Eq.\eqref{act1} in the main text.

Similarly, 
the four scalar amplitudes $S^\pm_\pm$ in Eq.\eqref{XX} are expressed in terms of one single 
amplitude ${\rm S}$ by the following relations: 
\be
S^{-}_{+}&=&\frac{m_{\rm H}^2-\epsilon}{m_{\rm H}^2}\, S^{+}_{-}\,,  \\
S^{+}_{-}&=&\frac{2}{m_{\rm H}^2}\,\left(\dot{S}^{-}_{-}+a^2 H S^{+}_{+}\right)    , \nonumber \\
S^{+}_{+}&=&-\frac{1}{Ha^2}\,\dot{S}^{-}_{-}     \nonumber   \\
&+&\frac{2Hm_{\rm H}^4 {\rm P}^2\,\dot{\rm S}+m_{\rm H}^6 {\rm P}^2\, {\rm S}-m_{\rm H}^4 (2{\rm P}^2+3 m_{\rm H}^2) S^{-}_{-}/a^2   }{
2H^2[3m_{\rm H}^4+2{\rm P}^2(2m_{\rm H}^2-\epsilon)]},  \nonumber 
\ee 
while 
\begin{widetext}
\be
S^{-}_{-}=a^2{\rm P}^2\frac{   -4H\mh^2 {\rm P}^2\,\dot{\rm S }+\left\{
2{\rm P}^2[(\mh^2-2H^2)(2\mh^2-\epsilon)-\mh^4]+3\mh^4(\mh^2-2H^2)
\right\}{\rm S} }{4{\rm P}^4(\mh^2-\epsilon)+6{\rm P}^2(2\mh^2-\epsilon)(\mh^2-2H^2)+9\mh^4(\mh^2-2H^2)}. 
\ee
\end{widetext}
It is crucial that all four  $S^\pm_\pm$ 
are expressed  in terms of one single amplitude ${\rm S}$ that fulfills the master 
equation obtainable  by varying the effective action given by  Eq.\eqref{act1} in the main text.
This shows that there is only one dynamical  DoF in the scalar sector. 
Therefore,  
together with the tensor and vector modes, the theory propagates five DoF. 

The effective action for the scalars contains the kinetic term $K$ expressed by Eq.(18) in the main text. One has 
\be
\frac{U}{K}=\frac{b_0+b_2 {\rm P}^2+b_4 {\rm P}^4+b_6 {\rm P}^6}{C\, (c_0+c_2{\rm P}^2+c_4{\rm P}^4) }
\ee
where 
\be
C&=&3\mh^4(\mh^2-2H^2),  \nonumber \\
c_0&=&9\mh^4(\mh^2-2H^2),  \nonumber \\
c_2&=&6(\mh^2-2H^2)(2\mh^2-\epsilon),  \nonumber \\
c_2&=&4(\mh^2-\epsilon), 
\ee
and also 
\be
b_0&=&27\,\mh^8 M^2_{\rm eff}(\mh^2-2H^2)^2,   \\
b_2&=&9\,\mh^4(\mh^2-2H^2)^2[4M^2_{\rm eff}(2\mh^2-\epsilon)-\mh^4  ],   \nonumber   \\
b_4&=&6\,\mh^4\left[8\mh^6-(20H^2+9\epsilon)\mh^4\right. \nonumber  \\
&+&\left.(8H^4+20H^2\epsilon+2H\dot{p}+\epsilon^2)\mh^2-4H^2(H\dot{p}+\epsilon^2)\right]       \nonumber \\
&+&12(M^2_{\rm eff}-\mh^2)\left[5\mh^8-6(2H^2+\epsilon)\mh^6\right.  \nonumber \\
&+& (8H^4+14H^2\epsilon+\epsilon^2)\mh^4 \nonumber \\
&-&\left.4H^2\epsilon(2H^2+\epsilon)\mh^2+4H^4\epsilon^2\right],  \nonumber \\
b_6&=&4(\mh^2-\epsilon)^2[4M^2_{\rm eff}(\mh^2-H^2)+\mh^2(\epsilon-2H^2-\mh^2)].  \nonumber 
\ee
Notice that these expressions contain $\dot{p}$ and hence the third derivative of the background scale factor $a(t)$. 
The ratio $c_s^2=b_6/(Cc_4)$ is the speed of sound expressed by Eq.\eqref{cs} in the main text.


%

\end{document}